\documentclass[aps,prl,reprint,groupedaddress]{revtex4-1}

\usepackage{graphicx}
\usepackage{amsmath}
\usepackage{amssymb}
\usepackage{mathrsfs}
\usepackage{color}
\usepackage{float}
\usepackage{physics}

\begin{document}

\title{Fisher-Information Efficient Metagratings for Transverse Displacement Metrology \\at the Measurement Independent Shot-Noise Limit}

\author{Zheng Xi}
\email[z.xi@tudelft.nl]{}
\affiliation{Optics Research Group, Delft University of Technology}
\author{Sander Konijnenberg}
\email[Previously in the Optics Research Group at TU-Delft, now at ASML Research]{}
\affiliation{Optics Research Group, Delft University of Technology}

\author{H.P. Urbach}
\affiliation{Optics Research Group, Delft University of Technology}

\date{\today}

\begin{abstract}
We derive the measurement independent precision bound for transverse displacement metrology using a classical source. Using tools from quantum estimation theory and metamaterial designs, we prove that the bound is achieved efficiently using optimized metagratings by detecting only a very small amount of photons used to probe the displacement, and we also reveal the direct link between the resonant property in the unit cell and the conditions to achieve the bound: one with the 0th dipole resonance and the other one with the anapole condition of the 1st dipole.   

\end{abstract} 
\maketitle

One crucial research area within optics is to use photons as  information carriers to develop sensitive metrology methods to measure a very small displacement \cite{hell1994breaking,rust2006sub,betzig2006imaging,hell2007far,bobroff1993recent,den2016optical,abbott2016observation,RN406,RN403,RN394,neugebauer2016polarization,xi2016accurate,PhysRevLett.121.193902,wei2018interferometric}. The problem is equivalent to a phase estimation problem of which the precision is fundamentally limited by the quantum statistical nature of light. Advances in tackling this problem have been pushing the limits of a number of key research fields such as interferometry\cite{abbott2016observation,RN406,RN403,RN394}, metrology in the semiconductor industry\cite{den2016optical} and super-resolution microscopy\cite{hell1994breaking,rust2006sub,betzig2006imaging,hell2007far}.   

Using classical sources, the measurement uncertainty is given by the shot-noise limit which scales as $1/\sqrt{N_\text{probe}}$ with $N_\text{probe}$ the number of photons used to probe the change in position per unit time. One might intuitively think that the best measurement scheme is to maximize the fraction of the number of probe photons that is actually detected since then no photon is lost and thus all the information about the displacement is preserved\cite{barnett2003ultimate}. But this scheme soon runs into a problem when one tries to push the precision of the measurement by using a large $N_\text{probe}$\cite{PhysRevA.98.043856}: while extremely high optical powers are available (for example, from several hundreds of watts to a few kilowatts at the beamsplitter inside an interferometer)\cite{aasi2013enhanced,abbott2016observation,abadie2011gravitational}, most of the fast photon-detectors do not allow direct exposure to such large powers\cite{bachor2004guide}. Thus an important question to be answered is: Can one recover all the information contained in a large number of $N_\text{probe}$ by detecting a small number of $N_\text{det}$ detected photons per unit time set by the limit of the detector?  In other words, how to develop an information efficient measurement scheme that compresses the full measured information into a small number of detected photons? 

This question has been answered in the development of large scale interferometers such as the one used in LIGO\cite{aasi2013enhanced,abbott2016observation, abadie2011gravitational}, which are used to measure extremely small longitudinal displacements. By setting the operating point of the interferometer near the so called \emph{dark fringe} of the Michelson interferometer, almost all the information about the longitudinal displacement is compressed into a small amount of detected photons with typical $N_\text{det}/N_\text{probe}$ less than $10^{-4}$, allowing one to make full use of the very large power inside the interferometer. %Moreover, advanced power recycling techniques allow one to recycle all the undetected light to increase the effective number of probe photons, leading to a significant improvement compared to the case that one detects all the photons directly\cite{drever1983gravitational,meers1988recycling,PhysRevLett.114.170801}. 
This scheme is mainly used for probing longitudinal displacements along the beam propagation direction. 
For transverse position measurements, i.e. measuring displacements that make an angle with the beam propagation direction, impressive progress has been made in super-resolution microscopy by centroid-fitting\cite{hell1994breaking,rust2006sub,betzig2006imaging,hell2007far,backlund2018fundamental,PhysRevLett.113.133902}. However, the small efficiency due to small optical cross section of tiny scatterers or fluorescent molecules largely limits the efficiency of those techniques\cite{taylor2013biological}. 

In this work, we combine the knowledge from the quantum estimation theory and modern tools of metamaterials to design metagratings of which the measured diffracted orders contain the highest Fisher Information (FI) about the transverse displacement. The proposed metagrating achieves the measurement independent shot-noise limit for $N_\text{probe}$ photons in spite of the fact that the number of detected photons $N_\text{det}$ satisfies $N_\text{det} \ll N_\text{probe}$. Using multipole scattering theory, the optimized designs are found to be closely related to the resonant properties of the unit cell of the metagrating, namely the 0th order dipole resonance and the 1st order anapole condition. 

The problem we consider is shown in Fig. 1(a). An arbitrary object M is placed inside an interference field formed by two plane waves which propagate in the xy plane at angles defined by the transverse wave vector component $\pm{k_x}$. The two plane waves have the same polarization and amplitude. They are out of phase. We set the object's original position as $x=0$ and then displace the object by $\delta x$. The question is how well can one measure the displacement $\delta x$ using classical light sources provided that one is free to choose any type of measurement object M and any type of unbiased estimator?

\begin{figure}
\includegraphics[width=\linewidth]{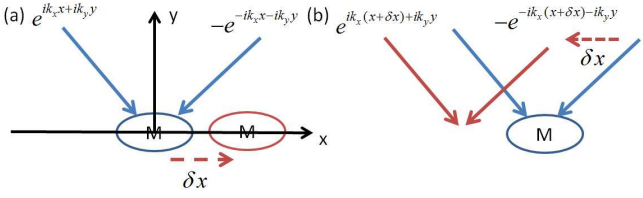}
    \caption{The problem of measuring a transverse displacement $\delta x$ using an object M. The illumination field is produced by two plane waves interference with an initial phase difference of $\pi$. (a) The object M is displaced by $\delta x$ from the origin is equivalent to (b) adding phase shifts of ${\pm k_x\delta x}$ to the two beams. }
    \label{fig:my_label}
\end{figure}

The Quantum Fisher Information (QFI) and the associated Quantum Cram\'er Rao Bound (QCRB) provide the answer to this question\cite{RN494}. The maximum amount of information encoded in the probe photons about the  parameter is given by the QFI. The QFI is optimized over all measurement schemes and all unbiased estimators. It can also be optimized over the initial quantum states used including N00N states etc, but in the following we restrict to coherent state since it is relatively easy to prepare a large $N_\text{probe}$ in such a state. From QFI, one can derive the QCRB which shows the measurement independent bound on the uncertainty of estimation.  

As shown in Fig. 1(b), the transverse displacement of the object M by $\delta x$ is equivalent to adding phase shifts of $\pm\delta\phi = {\pm k_x\delta x }$ to the two input beams. The QCRB for phase estimation is:
\begin{equation}
    \delta \phi^{\text{QCRB}}  = \frac{1}{\sqrt{{F_{Q\phi}}}},
\end{equation}
for which ${F_{Q\phi}}= 4\text{Var}(H)$ is the QFI for phase estimation expressed by the variance of the generator $H=a^\dagger a - b^\dagger b$ in the initial quantum state $\ket{\psi_0}=\ket{\frac{\alpha}{\sqrt{2}}}_a\ket{-\frac{\alpha}{\sqrt{2}}}_b$ with $a$ and $b$ being the annihilation operators of the left and right input modes\cite{taylor2016quantum}. $\ket{\alpha}$ is the input coherent state. It is easy to show that $F_{Q\phi}=4N_\text{probe}$ with $N_\text{probe}=|\alpha|^2$ being the total averaged number of probe photons per unit time in the two input modes.

Applying the chain rule, the QCRB using coherent state for determining the displacement $\delta x$ is:
\begin{equation}
    \delta x^\text{QCRB} = \frac{1}{\sqrt{k_x^2F_{Q\phi}}} =\frac{1}{2k_x\sqrt{N_\text{probe}}}.
\end{equation}
 Since we are using a classical coherent source and we refer to the QCRB shown in Eq. (2) as the measurement independent shot-noise limit. The proposed measurement scheme should be able to attain this limit with $N_\text{det}\ll N_\text{probe}$.

The measurement independent QFI and QCRB only provide a reference to check if the optimal bound is achieved, they do not provide detailed design guidelines on what kind of object M and the measurement scheme to choose. Related to the QFI is the concept of FI\cite{kay1993fundamentals}, which describes the maximum amount of information can be extracted about the  parameter of interest from a specific measurement scheme using any unbiased estimators. The main difference between QFI and FI is that the latter is dependent on the specific measurement chosen. Accordingly, the Cram\'er Rao Bound (CRB) gives the measurement dependent uncertainty bound. The advantages of using FI is that we can evaluate a specific class of measurement schemes characterized by some parameters. By maximizing FI by varying the parameters, one can derive the scheme for which the CRB is the smallest and if suitably chosen, the CRB eventually attains QCRB. In this way, the optimal measurement scheme can be identified.

\begin{figure}
\includegraphics[width=\linewidth]{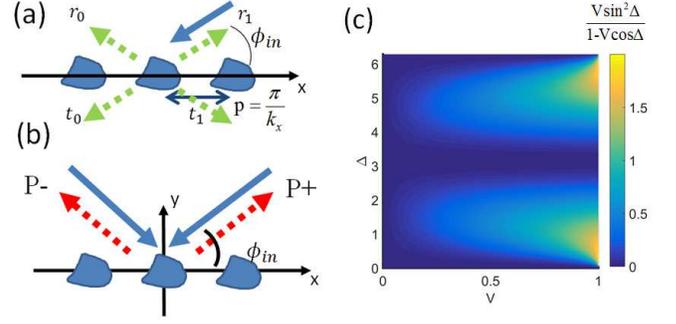}
    \caption{(a) A general grating considered. The pitch p is chosen to be $\pi/k_x$ with $k_x$ being the transverse wave vector of the incoming beam. For this configuration, there are four diffraction orders symmetrically emerging from the grating at angles defined by $\pm k_x$. (b) The grating is put inside the interference field and used as the measurement object M for transverse position metrology. Power difference between the two reflection orders are measured. (c) Plot of the function $\frac{V\text{sin}^2\Delta}{1-V\text{cos}\Delta}$}
    \label{fig:my_label}
\end{figure}

We consider a specific class of measurement objects: namely a grating made of a general unit cell structure as shown in Fig. 2(a). We have chosen the pitch to be $p = \pi/k_x$ and the incident angle $\phi_\text{in}$ to be smaller than $70^{\circ}$ to make sure that there are only four diffraction orders. The four orders have complex diffraction coefficients $r_0$, $r_1$, $t_0$ and $t_1$. When used as the measurement object, there are two reflected orders $P_+$ and $P_-$ parallel to the incident direction $\pm k_x$ shown in Fig. 2(b). These two reflected orders are each produced by the interference of the 0th and 1st reflection orders from the two incident beams. Due to this interference, the power in $P_+$ and $P_-$ are dependent on the displacement $\delta x$:
\begin{equation}
     \begin{split}
         {\mathscr{P}_{+}}(\delta x)=A\mathscr{P}[1-V\cos (2{{k}_{x}}\delta x-\Delta)],\\  
         {\mathscr{P}_{-}}(\delta x)=A\mathscr{P}[1-V\cos (2{{k}_{x}}\delta x+\Delta)],
     \end{split}
\end{equation}
for which $A=\frac{|r_{0}|^{2}+|r_{1}|^{2}}{2}$, $V=\frac{|2{{r}_{0}}{{r}_{1}}|}{|r_{0}|^{2}+|r_{1}|^{2}}$ is the visibility and $\Delta$ (in radians) is the phase difference between the two reflection coefficients $|r_0|$ and $|r_1|e^{i\Delta}$. By detecting the power difference in $P_+$ and $P_-$, the displacement $\delta x$ can be estimated. Under Poisson shot-noise which is directly linked to the coherent state used as mentioned, %the detected power $\mathscr{P_{+{\text{det}}}}$  and $\mathscr{P_{-{\text{det}}}}$ follow
%Poissonian distribution conditioned on the actual position x with  $\mathscr{P_{+{\text{det}}}}|x = \text{Poisson}(\mathscr{P_+};x)$, $\mathscr{P_{-{\text{det}}}}|x = \text{Poisson}(\mathscr{P_-};x)$. 
the FI contained in the detected reflection orders are:
\begin{equation}
      F_\text{det}(\delta x) = \frac{1}{N_\text{det}}\bigg(\frac{dN_\text{det}}{d\delta x}\bigg)^2,  
\end{equation}
with $N_\text{det}= [\mathscr{P}_{+}(\delta x)+{{\mathscr{P}}_{-}(\delta x)}]/ \hbar \omega$ is the averaged number of detected photons per unit time at position $\delta x$.
If one is interested in measuring very small displacements $\delta x \approx 0$, Eq. (4) becomes: 
\begin{equation}
%\begin{split}
  F_\text{det}(0) = 8Ak_x^2N_{\text{probe}}\frac{V\text{sin}^2\Delta}{1-V\text{cos}\Delta},\\
%\end{split}
\end{equation}
which is monotonically increasing with the visibility $V$. When $V=1$, it becomes:
\begin{equation}
%\begin{split}
  F_\text{det}(0) 
  =16Ak_x^2N_{\text{probe}}\text{cos}^2\frac{\Delta}{2}
  \leq 16|r_0|^2k_x^2N_{\text{probe}},
%\end{split}
\end{equation}
the maximum value of FI $F_\text{det}(0)$ is achieved for sufficiently small $\Delta$ with $V=1$ as shown in Fig. 2(c) (we want to avoid $\Delta=0, V=1$ as this point corresponds to no detected power). Referring to Eq. (3), this condition has a clear physical interpretation: it corresponds to a grating designed such that when displaced by a very small displacement $|\Delta/2k_x|$ left or right from the origin, one of the reflected orders $P_+$ or $P_-$ vanishes. 

The CRB in estimating $\delta x$ around the origin is thus:
\begin{equation}
%\begin{split}
    {{\delta x}^{\text{CRB}}}(0)=\frac{1}{\sqrt{F_\text{det}(0)}}
\geq \frac{1}{4|r_0|k_x\sqrt{N_\text{probe}}},
%\end{split}
\end{equation}
which is dependent on $|r_0|$. To achieve the measurement independent QCRB in Eq. (2), $|r_0|$ should be 1/2. 

Through this detailed comparison of the CRB and the QCRB, we can summarize the requirements for the optimal grating design: $V=1$ with $\Delta$ sufficiently small and $|r_0|=1/2$. These requirements can also be expressed as:
\begin{equation}
       |r_0|=\frac{1}{2},\quad r_1 = r_0e^{i\Delta} \quad \text{with  sufficiently small $\Delta$}.
\end{equation}

Under these requirements, the number of detected photons is given by $N_\text{det}=N_\text{probe}\text{sin}^2\Delta/2$ and is much smaller than $N_\text{probe}$. The fact that the QCRB is achieved proves that these detected photons contain all the information from the large $N_\text{probe}$. Thus the scheme is very Fisher-Information efficient. 

The properties of such a grating can be further revealed using multipole scattering theory commonly used
in the field of metamaterial design\cite{twersky1962scattering,twersky1956scatttering,yasumoto2005electromagnetic}. This theory relates the diffraction property to the resonances of one unit cell from which physical insights can be gained. Often the designed grating is referred to as metagrating emphasizing the importance of the unit cell resonances as well as the ability to tune these resonances with artificially designed unit cells\cite{epstein2017unveiling,PhysRevApplied.10.011002,PhysRevLett.119.067404}. In the following, we adapt this approach and show how the requirements mentioned can be realized.

Using multipole scattering theory, the diffraction efficiencies of the reflected and transmitted orders are:
\begin{equation}
\begin{split}
        r_\mu&=C_\mu(\phi_\mu)G(\phi_{\mu},\phi_\text{in})\\
        t_\mu&=\delta_{\mu 0}+C_\mu(2\pi-\phi_\mu)G(2\pi-\phi_{\mu},\phi_\text{in}),
\end{split}
\end{equation}
for which $C_\mu(\phi_\mu)=2/k_0p\text{sin}\phi_\mu$, $k_0$ is the wave number in the uniform surrounding material which we take to be air, $\phi_\mu$ is the diffraction angle of order $\mu$ measured from $+x$ axis. $G(\phi_\mu,\phi_\text{in})$ is the multiple scattering Green's function of the unit cell and can be expressed using multipole expansion:
\begin{equation}
{{G}}(\phi_{\mu},\phi_\text{in})=\sum\limits_{m=-\infty}^{+\infty }{{{A}_{m}(\phi_\text{in})}{{e}^{im\phi_\mu }}},
\label{}
\end{equation}
with ${{A}_{m}}$ corresponding to the multiple-scattered multipole coefficients and can be constructed from the isolated scattering coefficients $a_m$ for the same object inside the unit cell:
\begin{equation}
    A_m(\phi_\text{in}) = a_m\bigg[e^{im\phi_\text{in}}+\sum\limits_{n=-\infty}^{+\infty}A_n(\phi_\text{in})\mathscr{L}(n,m)\bigg],
\end{equation}
with $\mathscr{L}(n,m)$ being the lattice sum which takes fully account of the multiple scattering between different unit cells\cite{twersky1962scattering,twersky1956scatttering,yasumoto2005electromagnetic} and its convergence has been tested by including a sufficient number of terms.

We substitute Eq. (9) into the second requirement in Eq. (8): 
\begin{equation}
{{G}}({{\phi}_{1},\phi_\text{in}})={G}({{\phi }_{0}},\phi_\text{in}){{e}^{i\Delta}},
\label{}
\end{equation}
where $\phi_0 = \pi-\phi_\text{in}$ and $\phi_1=\phi_\text{in}$. For sufficiently small unit cell structure, one can keep only the first three terms ${{A}_{-1,0,1}}$ in Eq. (10) and Eq. (12) becomes:
\begin{equation}
     -{{A}_{0}}\sin\frac{\Delta}{2} +i{{A}_{1}}\cos ({{\phi }_{\text{in}}}-\frac{\Delta}{2})+i{{A}_{-1}}\cos ({{\phi }_{\text{in}}}+\frac{\Delta}{2})=0.  
\end{equation}
Hence, for sufficiently small $\Delta$:
\begin{equation}
\Delta \approx \frac{2i(A_1+A_{-1})\text{cos}\phi_\text{in}}{A_0}.
\end{equation}
It is interesting to look at the case when $A_0$ is the dominant term in the expansion. In this case $V\approx1$, $\Delta$ can be made small and the diffraction efficiencies in Eq. (9) can be approximated by $r_0 \approx {C_0(\phi_0)A_0(\phi_{in})}$, $ r_1 \approx  r_0$, $t_0 \approx 1 - r_0$ and $t_1 \approx -r_0 $, which, by energy conservation $|r_0|^2+|r_1|^2+|t_0|^2+|t_1|^2\approx1$, leads to $|r_0|\approx1/2$. All the design requirements imposed by Eq. (8) are fulfilled. Using this insight gained, we propose in the following two metagrating unit cell designs fulfilling the design requirements: the first one works at the $A_0$ resonance and the second works around the anapole condition of $a_{1}$ for which $A_{\pm 1}$ are very small.

In the following discussion, the wavelength is at 650nm and the incoming polarization state is TE polarization. The pitch is adjusted with the incident angle defined by $k_x$ with $p=\pi/k_x$. For a direct comparison, we calculate the FI in the detected photons using Eq. (5) with reflection coefficients $r_0$ and $r_1$ from Eq. (9), the obtained FI is normalized with QFI $4k^2_xN_\text{probe}$ from Eq. (2). 

\emph{Case I. Achieving shot-noise limit at the $A_0$ resonance.} In this case, the denominator of Eq. (14) is large due to the $A_0$ resonance. The unit cell is made of silicon cylinder of refractive index $n_\text{Si}=3.5$ with varying radius $R$ shown in Fig. 3(a). The incident angle  $\phi_\text{in}$ is $70^{\circ}$ to make the term $\text{cos}\phi_\text{in}$ in Eq. (14) small without introducing higher diffraction orders.
\begin{figure}
\includegraphics[width=\linewidth]{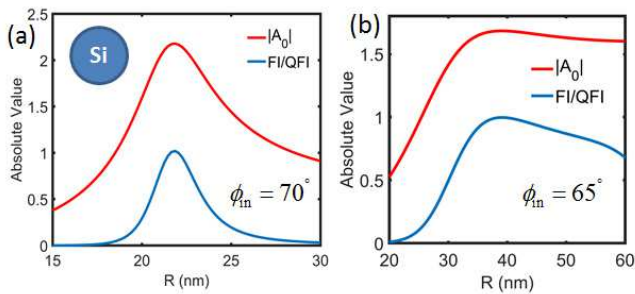}
    \caption{Normalized FI/QFI and $|A_0|$ for metagrating designs made of Si nanowire in the unit cell at (a) $\phi_\text{in}=70^{\circ}$ and (b) $\phi_\text{in}=65^{\circ}$. The FI is maximized at the $A_0$ resonance achieving QFI. }
    \label{fig:my_label}
\end{figure}
The FI reaches QFI at $A_0$ resonance around $R=22$nm with $\Delta \approx 4\times10^{-4}$. Notice that the FI is sharply peaked around $R = 22$nm, which is hard to achieve considering fabrication errors. %To have a more robust design, we take a look at Fig. 2(c). The $R=22$nm point corresponds to the point where $V=1$ and $\Delta \approx 4e^{-4}$, which has little tolerance in $V$. If there are some mismatches between $r_0$ and $r_1$ due to fabrication errors, the FI drops significantly as a function of $R$ as shown in Fig. 3(a).% 
The design can be made more robust by keeping $\Delta$ slightly larger which corresponds to move slightly from the bottom-right upwards in Fig. 2(c). %In this case, we move away from the bottom-right point shown in Fig. 2(c) slightly upwards, the FI still remains high but has a larger tolerance in $V$.
%For the metagrating design, this means we can slightly increase $\Delta$ by using a smaller angle.
In Fig. 3(b) we decrease the incident angle  $\phi_\text{in}=65^\circ$ and adjust the pitch accordingly to slightly increase $\Delta$. The optimum radius moves to $R=40$nm with $\Delta\approx 0.01$ and $V=1$. The FI remains high (FI/QFI$>$0.9) within a broader range $\Delta R = 10$nm. The $N_\text{det}/N_\text{probe}$ ratio is about $2.5\times10^{-5}$.

\emph{Case II. Achieving shot-noise limit around the $a_{1}$ anapole condition.} Another way to make $A_0$ dominant is to make the numerator containing $A_{\pm 1}$ small. This can be achieved around the anapole condition of $a_1$. Anapole is a radiationless state due to destructive interference of different modes in the far field\cite{miroshnichenko2015nonradiating,baryshnikova2019optical,gbur2001nonradiating,wei2016excitation}. At this condition, the single scattering multipole coefficient $a_m$ in the multipole expansion becomes approximately zero. If one can tune the structure such that the first dipole coefficient $a_1\approx0$, the effect of multiple scattering described by Eq. (11) becomes negligible and accordingly $A_{\pm 1} \approx 0$. According to Eq. (14), to make $\Delta$ real with finite $A_{\pm 1}$, an additional phase difference of $\pi/2$ is required. Therefore, we expect to see that the FI approaching QFI slightly away from the $a_1$ anapole condition.
\begin{figure}
\includegraphics[width=0.8\linewidth]{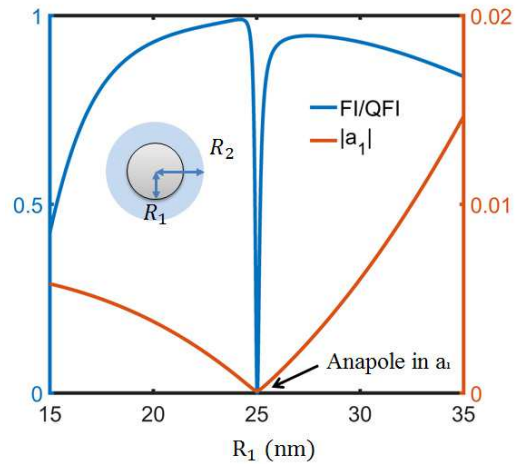}
     \caption{Normalized FI/QFI and $|a_1|$ for metagrating designs made of Ag-Core ${\text{SiO}_2}$-shell nanowire in the unit cell at $\phi_\text{in}=5^{\circ}$ with $R_2=50$nm and varying $R_1$. The FI is maximized around the $a_1$ anapole condition achieving QFI. }
    \label{fig:my_label}
\end{figure}
In Fig. 4, the unit cell made of a Ag-core $\text{SiO}_2$-shell cylinder with $n_\text{Ag}=0.052225 + 4.4094i$ and $n_{\text{SiO}_2}=1.5$ is used. The incident angle is chosen to be at $\phi_\text{in}=5^\circ$, which  makes $k_x$ large. The outer radius is kept as $R_2=50$nm with the varying inner radius $R_1$. The anapole condition is clearly seen at $R_1=25$nm which corresponds to vanishing $a_1$. FI comes close to QFI around $R=24$nm, slightly away from the anapole condition. Moreover, it remains relatively high within a certain range around the anapole condition. For this configuration, $\Delta \approx 0.04$ and $N_\text{det}/N_\text{probe}\approx 4\times10^{-4}$.

In conclusion, we have derived Fisher-Information efficient metagratings designs that reach the QFI for transverse displacement metrology with the number of detected photons several orders of magnitude smaller than the number of probe photons. Physical insight is obtained for the design using mulitpole expansion theory. It is found that around the 0th dipole resonance and the anapole condition of the 1st dipole, the FI comes very close to QFI which means the measurement independent shot-noise limit is achieved. The link between the fundamental precision limit under shot-noise and the multipole-based metagrating design will give new physical insights to the field of high precision metrology.

%The above derivation works for both TE and TM polarization. For TM polarization, the term ${{A}_{0}}$ corresponds to the multiple-scattered linearly polarized magnetic dipole in the array. Because of the broken symmetry, the ${{A}_{-1}}$ and ${{A}_{1}}$ terms are different when ${{\phi }_{\text{in}}}\ne {{90}^{\circ}}$. Since their scattered fields carry different topological charges $-1$ and $+1$ as shown in the insets of Fig. 2(a), they can be associated with the multiple-scattered left-rotating and right-rotating electric dipoles. For TE polarization, $A_0$ corresponds to the multiple-scattered linearly polarized electric dipole while ${{A}_{\pm1}}$ are the rotating magnetic dipoles.

%On the left-hand-side of Eq. (12) is the complex amplitude of reflection order $P_+$ at position $x=x_0$. To make this order vanish at $x_0$, there should be a proper balance of the different multipole components as indicated by Eq. (13). The term \emph{i} in front of ${{A}_{\pm1}}$ indicates a phase difference of $\pi/2$ is required. When one of the multipoles experiences a resonance, the phase around this resonance changes rapidly, by which this requirement can be fulfilled.

%\bibliography{ref}

\begin{thebibliography}{35}%
\makeatletter
\providecommand \@ifxundefined [1]{%
 \@ifx{#1\undefined}
}%
\providecommand \@ifnum [1]{%
 \ifnum #1\expandafter \@firstoftwo
 \else \expandafter \@secondoftwo
 \fi
}%
\providecommand \@ifx [1]{%
 \ifx #1\expandafter \@firstoftwo
 \else \expandafter \@secondoftwo
 \fi
}%
\providecommand \natexlab [1]{#1}%
\providecommand \enquote  [1]{``#1''}%
\providecommand \bibnamefont  [1]{#1}%
\providecommand \bibfnamefont [1]{#1}%
\providecommand \citenamefont [1]{#1}%
\providecommand \href@noop [0]{\@secondoftwo}%
\providecommand \href [0]{\begingroup \@sanitize@url \@href}%
\providecommand \@href[1]{\@@startlink{#1}\@@href}%
\providecommand \@@href[1]{\endgroup#1\@@endlink}%
\providecommand \@sanitize@url [0]{\catcode `\\12\catcode `\$12\catcode
  `\&12\catcode `\#12\catcode `\^12\catcode `\_12\catcode `\%12\relax}%
\providecommand \@@startlink[1]{}%
\providecommand \@@endlink[0]{}%
\providecommand \url  [0]{\begingroup\@sanitize@url \@url }%
\providecommand \@url [1]{\endgroup\@href {#1}{\urlprefix }}%
\providecommand \urlprefix  [0]{URL }%
\providecommand \Eprint [0]{\href }%
\providecommand \doibase [0]{http://dx.doi.org/}%
\providecommand \selectlanguage [0]{\@gobble}%
\providecommand \bibinfo  [0]{\@secondoftwo}%
\providecommand \bibfield  [0]{\@secondoftwo}%
\providecommand \translation [1]{[#1]}%
\providecommand \BibitemOpen [0]{}%
\providecommand \bibitemStop [0]{}%
\providecommand \bibitemNoStop [0]{.\EOS\space}%
\providecommand \EOS [0]{\spacefactor3000\relax}%
\providecommand \BibitemShut  [1]{\csname bibitem#1\endcsname}%
\let\auto@bib@innerbib\@empty
%</preamble>
\bibitem [{\citenamefont {Hell}\ and\ \citenamefont
  {Wichmann}(1994)}]{hell1994breaking}%
  \BibitemOpen
  \bibfield  {author} {\bibinfo {author} {\bibfnamefont {S.~W.}\ \bibnamefont
  {Hell}}\ and\ \bibinfo {author} {\bibfnamefont {J.}~\bibnamefont
  {Wichmann}},\ }\href@noop {} {\bibfield  {journal} {\bibinfo  {journal}
  {Optics Letters}\ }\textbf {\bibinfo {volume} {19}},\ \bibinfo {pages} {780}
  (\bibinfo {year} {1994})}\BibitemShut {NoStop}%
\bibitem [{\citenamefont {Rust}\ \emph {et~al.}(2006)\citenamefont {Rust},
  \citenamefont {Bates},\ and\ \citenamefont {Zhuang}}]{rust2006sub}%
  \BibitemOpen
  \bibfield  {author} {\bibinfo {author} {\bibfnamefont {M.~J.}\ \bibnamefont
  {Rust}}, \bibinfo {author} {\bibfnamefont {M.}~\bibnamefont {Bates}}, \ and\
  \bibinfo {author} {\bibfnamefont {X.}~\bibnamefont {Zhuang}},\ }\href@noop {}
  {\bibfield  {journal} {\bibinfo  {journal} {Nature Methods}\ }\textbf
  {\bibinfo {volume} {3}},\ \bibinfo {pages} {793} (\bibinfo {year}
  {2006})}\BibitemShut {NoStop}%
\bibitem [{\citenamefont {Betzig}\ \emph {et~al.}(2006)\citenamefont {Betzig},
  \citenamefont {Patterson}, \citenamefont {Sougrat}, \citenamefont
  {Lindwasser}, \citenamefont {Olenych}, \citenamefont {Bonifacino},
  \citenamefont {Davidson}, \citenamefont {Lippincott-Schwartz},\ and\
  \citenamefont {Hess}}]{betzig2006imaging}%
  \BibitemOpen
  \bibfield  {author} {\bibinfo {author} {\bibfnamefont {E.}~\bibnamefont
  {Betzig}}, \bibinfo {author} {\bibfnamefont {G.~H.}\ \bibnamefont
  {Patterson}}, \bibinfo {author} {\bibfnamefont {R.}~\bibnamefont {Sougrat}},
  \bibinfo {author} {\bibfnamefont {O.~W.}\ \bibnamefont {Lindwasser}},
  \bibinfo {author} {\bibfnamefont {S.}~\bibnamefont {Olenych}}, \bibinfo
  {author} {\bibfnamefont {J.~S.}\ \bibnamefont {Bonifacino}}, \bibinfo
  {author} {\bibfnamefont {M.~W.}\ \bibnamefont {Davidson}}, \bibinfo {author}
  {\bibfnamefont {J.}~\bibnamefont {Lippincott-Schwartz}}, \ and\ \bibinfo
  {author} {\bibfnamefont {H.~F.}\ \bibnamefont {Hess}},\ }\href@noop {}
  {\bibfield  {journal} {\bibinfo  {journal} {Science}\ }\textbf {\bibinfo
  {volume} {313}},\ \bibinfo {pages} {1642} (\bibinfo {year}
  {2006})}\BibitemShut {NoStop}%
\bibitem [{\citenamefont {Hell}(2007)}]{hell2007far}%
  \BibitemOpen
  \bibfield  {author} {\bibinfo {author} {\bibfnamefont {S.~W.}\ \bibnamefont
  {Hell}},\ }\href@noop {} {\bibfield  {journal} {\bibinfo  {journal}
  {Science}\ }\textbf {\bibinfo {volume} {316}},\ \bibinfo {pages} {1153}
  (\bibinfo {year} {2007})}\BibitemShut {NoStop}%
\bibitem [{\citenamefont {Bobroff}(1993)}]{bobroff1993recent}%
  \BibitemOpen
  \bibfield  {author} {\bibinfo {author} {\bibfnamefont {N.}~\bibnamefont
  {Bobroff}},\ }\href@noop {} {\bibfield  {journal} {\bibinfo  {journal}
  {Measurement Science and Technology}\ }\textbf {\bibinfo {volume} {4}},\
  \bibinfo {pages} {907} (\bibinfo {year} {1993})}\BibitemShut {NoStop}%
\bibitem [{\citenamefont {den Boef}(2016)}]{den2016optical}%
  \BibitemOpen
  \bibfield  {author} {\bibinfo {author} {\bibfnamefont {A.~J.}\ \bibnamefont
  {den Boef}},\ }\href@noop {} {\bibfield  {journal} {\bibinfo  {journal}
  {Surface Topography: Metrology and Properties}\ }\textbf {\bibinfo {volume}
  {4}},\ \bibinfo {pages} {023001} (\bibinfo {year} {2016})}\BibitemShut
  {NoStop}%
\bibitem [{\citenamefont {Abbott}\ \emph {et~al.}(2016)\citenamefont {Abbott},
  \citenamefont {Abbott}, \citenamefont {Abbott}, \citenamefont {Abernathy},
  \citenamefont {Acernese}, \citenamefont {Ackley}, \citenamefont {Adams},
  \citenamefont {Adams}, \citenamefont {Addesso}, \citenamefont {Adhikari}
  \emph {et~al.}}]{abbott2016observation}%
  \BibitemOpen
  \bibfield  {author} {\bibinfo {author} {\bibfnamefont {B.~P.}\ \bibnamefont
  {Abbott}}, \bibinfo {author} {\bibfnamefont {R.}~\bibnamefont {Abbott}},
  \bibinfo {author} {\bibfnamefont {T.}~\bibnamefont {Abbott}}, \bibinfo
  {author} {\bibfnamefont {M.}~\bibnamefont {Abernathy}}, \bibinfo {author}
  {\bibfnamefont {F.}~\bibnamefont {Acernese}}, \bibinfo {author}
  {\bibfnamefont {K.}~\bibnamefont {Ackley}}, \bibinfo {author} {\bibfnamefont
  {C.}~\bibnamefont {Adams}}, \bibinfo {author} {\bibfnamefont
  {T.}~\bibnamefont {Adams}}, \bibinfo {author} {\bibfnamefont
  {P.}~\bibnamefont {Addesso}}, \bibinfo {author} {\bibfnamefont
  {R.}~\bibnamefont {Adhikari}},  \emph {et~al.},\ }\href@noop {} {\bibfield
  {journal} {\bibinfo  {journal} {Phys. Rev. Lett.}\ }\textbf {\bibinfo
  {volume} {116}},\ \bibinfo {pages} {061102} (\bibinfo {year}
  {2016})}\BibitemShut {NoStop}%
\bibitem [{\citenamefont {Gardner}\ \emph {et~al.}(1993)\citenamefont
  {Gardner}, \citenamefont {Marable}, \citenamefont {Welch},\ and\
  \citenamefont {Thomas}}]{RN406}%
  \BibitemOpen
  \bibfield  {author} {\bibinfo {author} {\bibfnamefont {J.~R.}\ \bibnamefont
  {Gardner}}, \bibinfo {author} {\bibfnamefont {M.~L.}\ \bibnamefont
  {Marable}}, \bibinfo {author} {\bibfnamefont {G.~R.}\ \bibnamefont {Welch}},
  \ and\ \bibinfo {author} {\bibfnamefont {J.~E.}\ \bibnamefont {Thomas}},\
  }\href {\doibase 10.1103/PhysRevLett.70.3404} {\bibfield  {journal} {\bibinfo
   {journal} {Phys. Rev. Lett.}\ }\textbf {\bibinfo {volume} {70}},\ \bibinfo
  {pages} {3404} (\bibinfo {year} {1993})}\BibitemShut {NoStop}%
\bibitem [{\citenamefont {Giltner}\ \emph {et~al.}(1995)\citenamefont
  {Giltner}, \citenamefont {McGowan},\ and\ \citenamefont {Lee}}]{RN403}%
  \BibitemOpen
  \bibfield  {author} {\bibinfo {author} {\bibfnamefont {D.~M.}\ \bibnamefont
  {Giltner}}, \bibinfo {author} {\bibfnamefont {R.~W.}\ \bibnamefont
  {McGowan}}, \ and\ \bibinfo {author} {\bibfnamefont {S.~A.}\ \bibnamefont
  {Lee}},\ }\href {\doibase 10.1103/PhysRevLett.75.2638} {\bibfield  {journal}
  {\bibinfo  {journal} {Phys. Rev. Lett.}\ }\textbf {\bibinfo {volume} {75}},\
  \bibinfo {pages} {2638} (\bibinfo {year} {1995})}\BibitemShut {NoStop}%
\bibitem [{\citenamefont {Granata}\ \emph {et~al.}(2010)\citenamefont
  {Granata}, \citenamefont {Buy}, \citenamefont {Ward},\ and\ \citenamefont
  {Barsuglia}}]{RN394}%
  \BibitemOpen
  \bibfield  {author} {\bibinfo {author} {\bibfnamefont {M.}~\bibnamefont
  {Granata}}, \bibinfo {author} {\bibfnamefont {C.}~\bibnamefont {Buy}},
  \bibinfo {author} {\bibfnamefont {R.}~\bibnamefont {Ward}}, \ and\ \bibinfo
  {author} {\bibfnamefont {M.}~\bibnamefont {Barsuglia}},\ }\href {\doibase
  10.1103/PhysRevLett.105.231102} {\bibfield  {journal} {\bibinfo  {journal}
  {Phys. Rev. Lett.}\ }\textbf {\bibinfo {volume} {105}},\ \bibinfo {pages}
  {231102} (\bibinfo {year} {2010})}\BibitemShut {NoStop}%
\bibitem [{\citenamefont {Neugebauer}\ \emph {et~al.}(2016)\citenamefont
  {Neugebauer}, \citenamefont {Wo{\'z}niak}, \citenamefont {Bag}, \citenamefont
  {Leuchs},\ and\ \citenamefont {Banzer}}]{neugebauer2016polarization}%
  \BibitemOpen
  \bibfield  {author} {\bibinfo {author} {\bibfnamefont {M.}~\bibnamefont
  {Neugebauer}}, \bibinfo {author} {\bibfnamefont {P.}~\bibnamefont
  {Wo{\'z}niak}}, \bibinfo {author} {\bibfnamefont {A.}~\bibnamefont {Bag}},
  \bibinfo {author} {\bibfnamefont {G.}~\bibnamefont {Leuchs}}, \ and\ \bibinfo
  {author} {\bibfnamefont {P.}~\bibnamefont {Banzer}},\ }\href@noop {}
  {\bibfield  {journal} {\bibinfo  {journal} {Nature Communications}\ }\textbf
  {\bibinfo {volume} {7}},\ \bibinfo {pages} {11286} (\bibinfo {year}
  {2016})}\BibitemShut {NoStop}%
\bibitem [{\citenamefont {Xi}\ \emph {et~al.}(2016)\citenamefont {Xi},
  \citenamefont {Wei}, \citenamefont {Adam}, \citenamefont {Urbach},\ and\
  \citenamefont {Du}}]{xi2016accurate}%
  \BibitemOpen
  \bibfield  {author} {\bibinfo {author} {\bibfnamefont {Z.}~\bibnamefont
  {Xi}}, \bibinfo {author} {\bibfnamefont {L.}~\bibnamefont {Wei}}, \bibinfo
  {author} {\bibfnamefont {A.~J.~L.}\ \bibnamefont {Adam}}, \bibinfo {author}
  {\bibfnamefont {H.}~\bibnamefont {Urbach}}, \ and\ \bibinfo {author}
  {\bibfnamefont {L.}~\bibnamefont {Du}},\ }\href@noop {} {\bibfield  {journal}
  {\bibinfo  {journal} {Phys. Rev. Lett.}\ }\textbf {\bibinfo {volume} {117}},\
  \bibinfo {pages} {113903} (\bibinfo {year} {2016})}\BibitemShut {NoStop}%
\bibitem [{\citenamefont {Bag}\ \emph {et~al.}(2018)\citenamefont {Bag},
  \citenamefont {Neugebauer}, \citenamefont {Wo\ifmmode~\acute{z}\else
  \'{z}\fi{}niak}, \citenamefont {Leuchs},\ and\ \citenamefont
  {Banzer}}]{PhysRevLett.121.193902}%
  \BibitemOpen
  \bibfield  {author} {\bibinfo {author} {\bibfnamefont {A.}~\bibnamefont
  {Bag}}, \bibinfo {author} {\bibfnamefont {M.}~\bibnamefont {Neugebauer}},
  \bibinfo {author} {\bibfnamefont {P.}~\bibnamefont {Wo\ifmmode~\acute{z}\else
  \'{z}\fi{}niak}}, \bibinfo {author} {\bibfnamefont {G.}~\bibnamefont
  {Leuchs}}, \ and\ \bibinfo {author} {\bibfnamefont {P.}~\bibnamefont
  {Banzer}},\ }\href {\doibase 10.1103/PhysRevLett.121.193902} {\bibfield
  {journal} {\bibinfo  {journal} {Phys. Rev. Lett.}\ }\textbf {\bibinfo
  {volume} {121}},\ \bibinfo {pages} {193902} (\bibinfo {year}
  {2018})}\BibitemShut {NoStop}%
\bibitem [{\citenamefont {Wei}\ \emph {et~al.}(2018)\citenamefont {Wei},
  \citenamefont {Zayats},\ and\ \citenamefont
  {Rodr{\'\i}guez-Fortu{\~n}o}}]{wei2018interferometric}%
  \BibitemOpen
  \bibfield  {author} {\bibinfo {author} {\bibfnamefont {L.}~\bibnamefont
  {Wei}}, \bibinfo {author} {\bibfnamefont {A.~V.}\ \bibnamefont {Zayats}}, \
  and\ \bibinfo {author} {\bibfnamefont {F.~J.}\ \bibnamefont
  {Rodr{\'\i}guez-Fortu{\~n}o}},\ }\href@noop {} {\bibfield  {journal}
  {\bibinfo  {journal} {Phys. Rev. Lett.}\ }\textbf {\bibinfo {volume} {121}},\
  \bibinfo {pages} {193901} (\bibinfo {year} {2018})}\BibitemShut {NoStop}%
\bibitem [{\citenamefont {Barnett}\ \emph {et~al.}(2003)\citenamefont
  {Barnett}, \citenamefont {Fabre},\ and\ \citenamefont
  {Ma{\i}tre}}]{barnett2003ultimate}%
  \BibitemOpen
  \bibfield  {author} {\bibinfo {author} {\bibfnamefont {S.~M.}\ \bibnamefont
  {Barnett}}, \bibinfo {author} {\bibfnamefont {C.}~\bibnamefont {Fabre}}, \
  and\ \bibinfo {author} {\bibfnamefont {A.}~\bibnamefont {Ma{\i}tre}},\
  }\href@noop {} {\bibfield  {journal} {\bibinfo  {journal} {The European
  Physical Journal D-Atomic, Molecular, Optical and Plasma Physics}\ }\textbf
  {\bibinfo {volume} {22}},\ \bibinfo {pages} {513} (\bibinfo {year}
  {2003})}\BibitemShut {NoStop}%
\bibitem [{\citenamefont {Ataman}\ \emph {et~al.}(2018)\citenamefont {Ataman},
  \citenamefont {Preda},\ and\ \citenamefont {Ionicioiu}}]{PhysRevA.98.043856}%
  \BibitemOpen
  \bibfield  {author} {\bibinfo {author} {\bibfnamefont {S.}~\bibnamefont
  {Ataman}}, \bibinfo {author} {\bibfnamefont {A.}~\bibnamefont {Preda}}, \
  and\ \bibinfo {author} {\bibfnamefont {R.}~\bibnamefont {Ionicioiu}},\ }\href
  {\doibase 10.1103/PhysRevA.98.043856} {\bibfield  {journal} {\bibinfo
  {journal} {Phys. Rev. A}\ }\textbf {\bibinfo {volume} {98}},\ \bibinfo
  {pages} {043856} (\bibinfo {year} {2018})}\BibitemShut {NoStop}%
\bibitem [{\citenamefont {Aasi}\ \emph {et~al.}(2013)\citenamefont {Aasi},
  \citenamefont {Abadie}, \citenamefont {Abbott}, \citenamefont {Abbott},
  \citenamefont {Abbott}, \citenamefont {Abernathy}, \citenamefont {Adams},
  \citenamefont {Adams}, \citenamefont {Addesso}, \citenamefont {Adhikari}
  \emph {et~al.}}]{aasi2013enhanced}%
  \BibitemOpen
  \bibfield  {author} {\bibinfo {author} {\bibfnamefont {J.}~\bibnamefont
  {Aasi}}, \bibinfo {author} {\bibfnamefont {J.}~\bibnamefont {Abadie}},
  \bibinfo {author} {\bibfnamefont {B.}~\bibnamefont {Abbott}}, \bibinfo
  {author} {\bibfnamefont {R.}~\bibnamefont {Abbott}}, \bibinfo {author}
  {\bibfnamefont {T.}~\bibnamefont {Abbott}}, \bibinfo {author} {\bibfnamefont
  {M.}~\bibnamefont {Abernathy}}, \bibinfo {author} {\bibfnamefont
  {C.}~\bibnamefont {Adams}}, \bibinfo {author} {\bibfnamefont
  {T.}~\bibnamefont {Adams}}, \bibinfo {author} {\bibfnamefont
  {P.}~\bibnamefont {Addesso}}, \bibinfo {author} {\bibfnamefont
  {R.}~\bibnamefont {Adhikari}},  \emph {et~al.},\ }\href@noop {} {\bibfield
  {journal} {\bibinfo  {journal} {Nature Photonics}\ }\textbf {\bibinfo
  {volume} {7}},\ \bibinfo {pages} {613} (\bibinfo {year} {2013})}\BibitemShut
  {NoStop}%
\bibitem [{\citenamefont {Abadie}\ \emph {et~al.}(2011)\citenamefont {Abadie},
  \citenamefont {Abbott}, \citenamefont {Abbott}, \citenamefont {Abbott},
  \citenamefont {Abernathy}, \citenamefont {Adams}, \citenamefont {Adhikari},
  \citenamefont {Affeldt}, \citenamefont {Allen}, \citenamefont {Allen} \emph
  {et~al.}}]{abadie2011gravitational}%
  \BibitemOpen
  \bibfield  {author} {\bibinfo {author} {\bibfnamefont {J.}~\bibnamefont
  {Abadie}}, \bibinfo {author} {\bibfnamefont {B.~P.}\ \bibnamefont {Abbott}},
  \bibinfo {author} {\bibfnamefont {R.}~\bibnamefont {Abbott}}, \bibinfo
  {author} {\bibfnamefont {T.~D.}\ \bibnamefont {Abbott}}, \bibinfo {author}
  {\bibfnamefont {M.}~\bibnamefont {Abernathy}}, \bibinfo {author}
  {\bibfnamefont {C.}~\bibnamefont {Adams}}, \bibinfo {author} {\bibfnamefont
  {R.}~\bibnamefont {Adhikari}}, \bibinfo {author} {\bibfnamefont
  {C.}~\bibnamefont {Affeldt}}, \bibinfo {author} {\bibfnamefont
  {B.}~\bibnamefont {Allen}}, \bibinfo {author} {\bibfnamefont
  {G.}~\bibnamefont {Allen}},  \emph {et~al.},\ }\href@noop {} {\bibfield
  {journal} {\bibinfo  {journal} {Nature Physics}\ }\textbf {\bibinfo {volume}
  {7}},\ \bibinfo {pages} {962} (\bibinfo {year} {2011})}\BibitemShut {NoStop}%
\bibitem [{\citenamefont {Bachor}\ \emph {et~al.}(2004)\citenamefont {Bachor},
  \citenamefont {Ralph}, \citenamefont {Lucia},\ and\ \citenamefont
  {Ralph}}]{bachor2004guide}%
  \BibitemOpen
  \bibfield  {author} {\bibinfo {author} {\bibfnamefont {H.-A.}\ \bibnamefont
  {Bachor}}, \bibinfo {author} {\bibfnamefont {T.~C.}\ \bibnamefont {Ralph}},
  \bibinfo {author} {\bibfnamefont {S.}~\bibnamefont {Lucia}}, \ and\ \bibinfo
  {author} {\bibfnamefont {T.~C.}\ \bibnamefont {Ralph}},\ }\href@noop {}
  {\emph {\bibinfo {title} {A guide to experiments in quantum optics}}},\
  Vol.~\bibinfo {volume} {1}\ (\bibinfo  {publisher} {Wiley Online Library},\
  \bibinfo {year} {2004})\BibitemShut {NoStop}%
\bibitem [{\citenamefont {Backlund}\ \emph {et~al.}(2018)\citenamefont
  {Backlund}, \citenamefont {Shechtman},\ and\ \citenamefont
  {Walsworth}}]{backlund2018fundamental}%
  \BibitemOpen
  \bibfield  {author} {\bibinfo {author} {\bibfnamefont {M.~P.}\ \bibnamefont
  {Backlund}}, \bibinfo {author} {\bibfnamefont {Y.}~\bibnamefont {Shechtman}},
  \ and\ \bibinfo {author} {\bibfnamefont {R.~L.}\ \bibnamefont {Walsworth}},\
  }\href@noop {} {\bibfield  {journal} {\bibinfo  {journal} {Physical review
  letters}\ }\textbf {\bibinfo {volume} {121}},\ \bibinfo {pages} {023904}
  (\bibinfo {year} {2018})}\BibitemShut {NoStop}%
\bibitem [{\citenamefont {Shechtman}\ \emph {et~al.}(2014)\citenamefont
  {Shechtman}, \citenamefont {Sahl}, \citenamefont {Backer},\ and\
  \citenamefont {Moerner}}]{PhysRevLett.113.133902}%
  \BibitemOpen
  \bibfield  {author} {\bibinfo {author} {\bibfnamefont {Y.}~\bibnamefont
  {Shechtman}}, \bibinfo {author} {\bibfnamefont {S.~J.}\ \bibnamefont {Sahl}},
  \bibinfo {author} {\bibfnamefont {A.~S.}\ \bibnamefont {Backer}}, \ and\
  \bibinfo {author} {\bibfnamefont {W.~E.}\ \bibnamefont {Moerner}},\ }\href
  {\doibase 10.1103/PhysRevLett.113.133902} {\bibfield  {journal} {\bibinfo
  {journal} {Phys. Rev. Lett.}\ }\textbf {\bibinfo {volume} {113}},\ \bibinfo
  {pages} {133902} (\bibinfo {year} {2014})}\BibitemShut {NoStop}%
\bibitem [{\citenamefont {Taylor}\ \emph {et~al.}(2013)\citenamefont {Taylor},
  \citenamefont {Janousek}, \citenamefont {Daria}, \citenamefont {Knittel},
  \citenamefont {Hage}, \citenamefont {Bachor},\ and\ \citenamefont
  {Bowen}}]{taylor2013biological}%
  \BibitemOpen
  \bibfield  {author} {\bibinfo {author} {\bibfnamefont {M.~A.}\ \bibnamefont
  {Taylor}}, \bibinfo {author} {\bibfnamefont {J.}~\bibnamefont {Janousek}},
  \bibinfo {author} {\bibfnamefont {V.}~\bibnamefont {Daria}}, \bibinfo
  {author} {\bibfnamefont {J.}~\bibnamefont {Knittel}}, \bibinfo {author}
  {\bibfnamefont {B.}~\bibnamefont {Hage}}, \bibinfo {author} {\bibfnamefont
  {H.-A.}\ \bibnamefont {Bachor}}, \ and\ \bibinfo {author} {\bibfnamefont
  {W.~P.}\ \bibnamefont {Bowen}},\ }\href@noop {} {\bibfield  {journal}
  {\bibinfo  {journal} {Nature Photonics}\ }\textbf {\bibinfo {volume} {7}},\
  \bibinfo {pages} {229} (\bibinfo {year} {2013})}\BibitemShut {NoStop}%
\bibitem [{\citenamefont {Helstrom}(1976)}]{RN494}%
  \BibitemOpen
  \bibfield  {author} {\bibinfo {author} {\bibfnamefont {C.~W.}\ \bibnamefont
  {Helstrom}},\ }\href {http://www.gbv.de/dms/hbz/toc/ht001227764.pdf} {\emph
  {\bibinfo {title} {Quantum detection and estimation theory}}},\ Mathematics
  in Science and Engineering\ (\bibinfo  {publisher} {Academic Press},\
  \bibinfo {address} {New York},\ \bibinfo {year} {1976})\BibitemShut {NoStop}%
\bibitem [{\citenamefont {Taylor}\ and\ \citenamefont
  {Bowen}(2016)}]{taylor2016quantum}%
  \BibitemOpen
  \bibfield  {author} {\bibinfo {author} {\bibfnamefont {M.~A.}\ \bibnamefont
  {Taylor}}\ and\ \bibinfo {author} {\bibfnamefont {W.~P.}\ \bibnamefont
  {Bowen}},\ }\href@noop {} {\bibfield  {journal} {\bibinfo  {journal} {Physics
  Reports}\ }\textbf {\bibinfo {volume} {615}},\ \bibinfo {pages} {1} (\bibinfo
  {year} {2016})}\BibitemShut {NoStop}%
\bibitem [{\citenamefont {Kay}(1993)}]{kay1993fundamentals}%
  \BibitemOpen
  \bibfield  {author} {\bibinfo {author} {\bibfnamefont {S.~M.}\ \bibnamefont
  {Kay}},\ }\href@noop {} {\emph {\bibinfo {title} {Fundamentals of statistical
  signal processing}}}\ (\bibinfo  {publisher} {Prentice Hall PTR},\ \bibinfo
  {year} {1993})\BibitemShut {NoStop}%
\bibitem [{\citenamefont {Twersky}(1962)}]{twersky1962scattering}%
  \BibitemOpen
  \bibfield  {author} {\bibinfo {author} {\bibfnamefont {V.}~\bibnamefont
  {Twersky}},\ }\href@noop {} {\bibfield  {journal} {\bibinfo  {journal} {IRE
  Transactions on Antennas and Propagation}\ }\textbf {\bibinfo {volume}
  {10}},\ \bibinfo {pages} {737} (\bibinfo {year} {1962})}\BibitemShut
  {NoStop}%
\bibitem [{\citenamefont {Twersky}(1956)}]{twersky1956scatttering}%
  \BibitemOpen
  \bibfield  {author} {\bibinfo {author} {\bibfnamefont {V.}~\bibnamefont
  {Twersky}},\ }\href@noop {} {\bibfield  {journal} {\bibinfo  {journal} {IRE
  Transactions on Antennas and Propagation}\ }\textbf {\bibinfo {volume} {4}},\
  \bibinfo {pages} {330} (\bibinfo {year} {1956})}\BibitemShut {NoStop}%
\bibitem [{\citenamefont {Yasumoto}(2005)}]{yasumoto2005electromagnetic}%
  \BibitemOpen
  \bibfield  {author} {\bibinfo {author} {\bibfnamefont {K.}~\bibnamefont
  {Yasumoto}},\ }\href@noop {} {\emph {\bibinfo {title} {Electromagnetic theory
  and applications for photonic crystals}}}\ (\bibinfo  {publisher} {CRC
  press},\ \bibinfo {year} {2005})\BibitemShut {NoStop}%
\bibitem [{\citenamefont {Epstein}\ and\ \citenamefont
  {Rabinovich}(2017)}]{epstein2017unveiling}%
  \BibitemOpen
  \bibfield  {author} {\bibinfo {author} {\bibfnamefont {A.}~\bibnamefont
  {Epstein}}\ and\ \bibinfo {author} {\bibfnamefont {O.}~\bibnamefont
  {Rabinovich}},\ }\href@noop {} {\bibfield  {journal} {\bibinfo  {journal}
  {Physical Review Applied}\ }\textbf {\bibinfo {volume} {8}},\ \bibinfo
  {pages} {054037} (\bibinfo {year} {2017})}\BibitemShut {NoStop}%
\bibitem [{\citenamefont {Popov}\ \emph {et~al.}(2018)\citenamefont {Popov},
  \citenamefont {Boust},\ and\ \citenamefont
  {Burokur}}]{PhysRevApplied.10.011002}%
  \BibitemOpen
  \bibfield  {author} {\bibinfo {author} {\bibfnamefont {V.}~\bibnamefont
  {Popov}}, \bibinfo {author} {\bibfnamefont {F.}~\bibnamefont {Boust}}, \ and\
  \bibinfo {author} {\bibfnamefont {S.~N.}\ \bibnamefont {Burokur}},\ }\href
  {\doibase 10.1103/PhysRevApplied.10.011002} {\bibfield  {journal} {\bibinfo
  {journal} {Phys. Rev. Applied}\ }\textbf {\bibinfo {volume} {10}},\ \bibinfo
  {pages} {011002} (\bibinfo {year} {2018})}\BibitemShut {NoStop}%
\bibitem [{\citenamefont {Ra'di}\ \emph {et~al.}(2017)\citenamefont {Ra'di},
  \citenamefont {Sounas},\ and\ \citenamefont
  {Al\`u}}]{PhysRevLett.119.067404}%
  \BibitemOpen
  \bibfield  {author} {\bibinfo {author} {\bibfnamefont {Y.}~\bibnamefont
  {Ra'di}}, \bibinfo {author} {\bibfnamefont {D.~L.}\ \bibnamefont {Sounas}}, \
  and\ \bibinfo {author} {\bibfnamefont {A.}~\bibnamefont {Al\`u}},\ }\href
  {\doibase 10.1103/PhysRevLett.119.067404} {\bibfield  {journal} {\bibinfo
  {journal} {Phys. Rev. Lett.}\ }\textbf {\bibinfo {volume} {119}},\ \bibinfo
  {pages} {067404} (\bibinfo {year} {2017})}\BibitemShut {NoStop}%
\bibitem [{\citenamefont {Miroshnichenko}\ \emph {et~al.}(2015)\citenamefont
  {Miroshnichenko}, \citenamefont {Evlyukhin}, \citenamefont {Yu},
  \citenamefont {Bakker}, \citenamefont {Chipouline}, \citenamefont
  {Kuznetsov}, \citenamefont {Luk’yanchuk}, \citenamefont {Chichkov},\ and\
  \citenamefont {Kivshar}}]{miroshnichenko2015nonradiating}%
  \BibitemOpen
  \bibfield  {author} {\bibinfo {author} {\bibfnamefont {A.~E.}\ \bibnamefont
  {Miroshnichenko}}, \bibinfo {author} {\bibfnamefont {A.~B.}\ \bibnamefont
  {Evlyukhin}}, \bibinfo {author} {\bibfnamefont {Y.~F.}\ \bibnamefont {Yu}},
  \bibinfo {author} {\bibfnamefont {R.~M.}\ \bibnamefont {Bakker}}, \bibinfo
  {author} {\bibfnamefont {A.}~\bibnamefont {Chipouline}}, \bibinfo {author}
  {\bibfnamefont {A.~I.}\ \bibnamefont {Kuznetsov}}, \bibinfo {author}
  {\bibfnamefont {B.}~\bibnamefont {Luk’yanchuk}}, \bibinfo {author}
  {\bibfnamefont {B.~N.}\ \bibnamefont {Chichkov}}, \ and\ \bibinfo {author}
  {\bibfnamefont {Y.~S.}\ \bibnamefont {Kivshar}},\ }\href@noop {} {\bibfield
  {journal} {\bibinfo  {journal} {Nature communications}\ }\textbf {\bibinfo
  {volume} {6}},\ \bibinfo {pages} {8069} (\bibinfo {year} {2015})}\BibitemShut
  {NoStop}%
\bibitem [{\citenamefont {Baryshnikova}\ \emph {et~al.}(2019)\citenamefont
  {Baryshnikova}, \citenamefont {Smirnova}, \citenamefont {Luk'yanchuk},\ and\
  \citenamefont {Kivshar}}]{baryshnikova2019optical}%
  \BibitemOpen
  \bibfield  {author} {\bibinfo {author} {\bibfnamefont {K.~V.}\ \bibnamefont
  {Baryshnikova}}, \bibinfo {author} {\bibfnamefont {D.~A.}\ \bibnamefont
  {Smirnova}}, \bibinfo {author} {\bibfnamefont {B.~S.}\ \bibnamefont
  {Luk'yanchuk}}, \ and\ \bibinfo {author} {\bibfnamefont {Y.~S.}\ \bibnamefont
  {Kivshar}},\ }\href@noop {} {\bibfield  {journal} {\bibinfo  {journal}
  {Advanced Optical Materials}\ ,\ \bibinfo {pages} {1801350}} (\bibinfo {year}
  {2019})}\BibitemShut {NoStop}%
\bibitem [{\citenamefont {Gbur}(2001)}]{gbur2001nonradiating}%
  \BibitemOpen
  \bibfield  {author} {\bibinfo {author} {\bibfnamefont {G.}~\bibnamefont
  {Gbur}},\ }\emph {\bibinfo {title} {Nonradiating sources and the inverse
  source problem}},\ \href@noop {} {Ph.D. thesis},\ \bibinfo  {school}
  {University of Rochester. Dept. of Physics and Astronomy} (\bibinfo {year}
  {2001})\BibitemShut {NoStop}%
\bibitem [{\citenamefont {Wei}\ \emph {et~al.}(2016)\citenamefont {Wei},
  \citenamefont {Xi}, \citenamefont {Bhattacharya},\ and\ \citenamefont
  {Urbach}}]{wei2016excitation}%
  \BibitemOpen
  \bibfield  {author} {\bibinfo {author} {\bibfnamefont {L.}~\bibnamefont
  {Wei}}, \bibinfo {author} {\bibfnamefont {Z.}~\bibnamefont {Xi}}, \bibinfo
  {author} {\bibfnamefont {N.}~\bibnamefont {Bhattacharya}}, \ and\ \bibinfo
  {author} {\bibfnamefont {H.~P.}\ \bibnamefont {Urbach}},\ }\href@noop {}
  {\bibfield  {journal} {\bibinfo  {journal} {Optica}\ }\textbf {\bibinfo
  {volume} {3}},\ \bibinfo {pages} {799} (\bibinfo {year} {2016})}\BibitemShut
  {NoStop}%
\end{thebibliography}


\end{document}